\documentclass[twocolumn, showpacs]{revtex4}
\usepackage{CJK}
\usepackage{graphics}
\usepackage{amsmath}
\usepackage{mathrsfs}
\usepackage{amsfonts}
\usepackage{overpic}
\usepackage{hyperref}
\usepackage{rotating}
\usepackage{amssymb}
\newcommand{\ket}[1]{|#1\rangle}
\newcommand{\bra}[1]{\langle #1|}
\newcommand{\inp}[2]{\langle #1 | #2\rangle}
\newcommand{\ex}[1]{\ensuremath{\left\langle{#1}\right\rangle}}
\newcommand{\exs}[1]{\ensuremath{\langle{#1}\rangle}}

\begin{document}
\begin{CJK*}{GB}{gbsn}
\title{Overlap with the Separable State
and Phase Transition in the Dicke Model: Zero and Finite Temperature
}
\author{H. T. Cui(´Þº£ÌÎ)}
\email{cuiht@aynu.edu.cn} \affiliation{School of Physics and
Electric Engineering, Anyang Normal University, Anyang 455000,
China}
\date{\today}
\begin{abstract}
Overlap with the separable state is introduced in this paper for the
purpose of characterizing the overall correlation in many-body
systems. This definition has clear geometric and physical meaning,
and moreover can be considered as the generalization of the
concept-Anderson Orthogonality Catastrophe. As an exemplification,
it is used to mark the phase transition in the Dicke model for zero
and finite temperature. And our discussion shows that it can
faithfully reflect the phase transition properties of this model
whether for zero or finite temperature. Furthermore the overlap for
ground state also indicates the appearance of multipartite
entanglement in Dicke model.
\end{abstract}
 \maketitle
\end{CJK*}


\section{introduction}
Correlation in condensed matter systems predominates the
understanding of many-body effects. Fundamentally one can define
different correlation functions for describing the unusual
connections in many-body systems. For instance it is general to
introduce the order parameter for the description of phase
transitions induced by local perturbation, and furthermore to
classify the diverse phase transitions by scaling the singularity of
correlation functions with the universal critical exponents. That is
so called Landau-Ginzburg-Wilson(LGW) paradigm\cite{senthil04}.
However the situation becomes different for the strongly correlated
electronic systems. The quantum Hall effect appearing in
two-dimensional electron gas with high magnetic field shows the
distinct features not captured by LGW paradigm. Instead topological
order consequently is defined to describe the underlying symmetry in
quantum Hall systems, which is distinct from the notion of
spontaneously broken symmetry\cite{rw}.

Recently the extensive researches of quantum entanglement in
condensed mattered systems show the potentiality that quantum
entanglement would act the universal description for many-body
effects\cite{afov07, ecp08}. Especially some general conclusions
have been obtained about the connection between quantum entanglement
and quantum phase transition in many-body systems. The concurrence,
a measurement of two-party entanglement, has been shown to behave
singularly at the critical point of one-dimensional spin-half XY
model, and the critical exponents can also be obtained by scaling
this singularity\cite{oaff02, on02}. Furthermore the block
entanglement entropy has been shown to display the logarithmical
divergency with the block size at the critical points, and the
scaling factor is directly related to the central charge of the
conformal field theory\cite{cc04}. Moreover the universal area law
for the entanglement entropy has also been constructed exactly in
one dimensional spin-chain systems\cite{ecp08}, and the similar
behavior for single-copy entanglement is also founded\cite{sc}.
Recently the entanglement spectrum has been defined to obtain the
general information about the many-body systems\cite{lh08, cl08,
yq10}. As for quantum Hall systems, it is shown that the scaling
behavior of entanglement entropy is directly related to the quantum
number, which is used to characterize the topological
order\cite{topo06}. And entanglement spectrum can also be used to
detect the non-local features of quantum Hall systems\cite{lh08}.

Although these important progresses have been made, there are a few
exceptions that lead to the suspicion of the validity of quantum
entanglement as an universal description for many-body effects.
Entanglement entropy sometimes provides ambiguous information about
the phase transitions in higher dimensional many-body
systems\cite{ecp08}. Even for one dimensional systems, it cannot
present the complete information in some situations. As an example,
the recent studies show that the block entanglement entropy for the
Valence-Bond-Solid(VBS) state of integer spin seems unsensible to
the degeneracy manifested by the underlying topological symmetry and
also does not display the dependence on the parity of spin number
$s$, which however both can be manifested clearly by introducing
string order parameter\cite{katsura}. As for quantum Hall systems,
the entanglement entropy and entanglement spectrum have also been
shown the limited ability of identifying the topological
orders\cite{yq10}.

In my point, this defect would attribute to the trace-out of the
superfluous degrees of freedom when one obtains the reduced density
matrix. And some information for the global features in many-body
systems is inevitably lost. This point has been exemplified in a
recent paper of our group\cite{cwy10}, in which the geometric
entanglement(GE) as a measurement of multipartite entanglement is
calculated for VBS state. The interesting result in this paper is
that GE displays two different scaling behaviors dependent on the
parity of spin number $s$, and the global GE is divergent linearly
with the particle number.

Through this short introduction, it seems promising to measure
multipartite entanglement in order to obtain the complete
information for many-body effects. Recently some efforts have been
made in this direction. The connection of multipartite entanglement
and quantum phase transition has been discussed in some special
models\cite{cwy10, wei05, GE, orus08}. However the crucial obstacle
for further development is the absence of the unified understanding
of the multipartite entanglement\cite{pv07, bgh10}. Whereas the
maximally entangled state can be defined unambiguously for bipartite
systems, what is the maximally entangled state for multipartite
systems is unclear until now\cite{bgh10}. Fortunately it is well
accepted that the fully separable state can be defined as
\begin{equation}\label{sep}
\rho^{sep}=\sum_i p_i
\rho^{(i)}_1\otimes\rho^{(i)}_2\otimes\cdots\rho^{(i)}_N,
\end{equation}
where $N$ is the particle number, and $p_i$ denotes the common
probability with which the single-particle state
$\rho_n^{(i)}(n=1,2,\cdots,N)$ happens. With respect of this point,
geometric entanglement(GE) is introduced first by Shimony for pure
bipartite state\cite{shimony95} and generalized to the multipartite
case by Carteret {\it et al.}\cite{chs00}, Barnum and Linden
\cite{bl01}, Wei and Goldbart \cite{wg03}, and to the mixed state by
Cao and Wang \cite{cw07}. GE is a genuine multipartite entanglement
measurement. The main idea of GE is to minimize the distance $D$
between the state $\ket{\Psi}$ to be measured and the fully
separable state $\ket{\Phi}$ in Hilbert space,
\begin{equation}\label{D}
D=\min_{\{\ket{\Phi}\}}\{\|\ket{\Psi}-\ket{\Phi}\|^2\}.
\end{equation}
For the normalized $\ket{\Psi}$ and $\ket{\Phi}$, the evaluation of
$D$ is reduced to find the maximal overlap\cite{wg03}
\begin{equation}\label{purelap}
\Lambda(\Psi)=\max_{\{\ket{\Phi}\}}|\inp{\Phi}{\Psi}|.
\end{equation}
Geometrically $\Lambda(\ket{\Psi})$ depicts the overlap angle
between the vectors $\ket{\Psi}$ and $\ket{\Phi}$ in Hilbert space.
Then the larger $\Lambda(\ket{\Psi})$ is, the shorter is the
distance and the less entangled is $\ket{\Psi}$. But the optimum is
in general a forbidden task, not spoken for mixed state, and the
analytical results can be obtained only for some very special
cases\cite{GE, cw07, orus08}. Recently many efforts are devoted to
the reduction of the optimum and some interesting results are
obtained\cite{wei09}.

Given this difficulty, we introduce another different quantity in
this paper to capture the overall correlation in condensed matter
systems, i.e. the overlap with a special fully separable state. The
starting point is still to find the minimal distance between the
state to be measured and a special fully separable state defined in
the next section. In contrast to GE the optimum can be reduced by
utilizing the geometric property of the overlap, and this overlap
has very clear physical meaning, whether for pure or mixed state. In
Sec.II, the definition of this overlap is introduced, and the
differences with several known similar definitions are clarified.
Furthermore we point out that our definition is connected intimately
with the concept of Anderson Orthogonality
Catastrophe(AOC)\cite{anderson67, mahan}. As an illustration of the
validity of our definition, the collective phase transition appeared
in Dicke model is discussed by this quantity in Sec.III.
Multipartite entanglement in this model is also studied for
displaying the potential connection between this overlap and
multipartite entanglement. Finally conclusions and further
discussion are presented in Sec.IV.

\section{overlap with fully separable state}

Similar to the introduction of GE, our starting point is also to
find the minimal distance $D$ between the fully separable state
$\rho^{sep}$ and the state $\rho$ to be measured.
\begin{equation}\label{D}
D=\min_{\{\rho^{sep}\}}\{\|\rho-\rho^{sep}\|^2\}.
\end{equation}
Generally this minimal distance is still decided mainly by the
maximal overlap
\begin{equation}\label{overlap}
\Lambda=\max_{\{\rho^{sep}\}}\text{Tr}[\rho \rho^{sep}].
\end{equation}

The density matrix can also be written as the Bloch form
\begin{equation}
\rho=(I+\sum_{i=1}^{d^2-1} r_i\lambda_i)/d,
\end{equation}
where $d$ denotes the dimension, $\lambda_i$ is the generator of
$SU(d)$ group and $\{r_i\}$ is so called Bloch vector\cite{he81}.
Thus $\Lambda$ has clear geometric meaning which depicts the minimal
overlap angle $\theta$ between the Bloch vectors $\{r_i\}$ and
$\{r_i\}_{sep}$ in the Bloch-vector space, i.e.,
\begin{equation}
\max_{\{\rho^{sep}\}}\text{Tr}[\rho \rho^{sep}]=\frac{1}{d}(1 +
|\{r_i\}||\{r_i\}_{sep}| \cos[\min_{\{\theta\}} \theta]).
\end{equation}
Two limit cases are beneficial to the understanding of the physical
meaning of $\theta$. For $\cos\theta=1$, the overlap is maximal and
$\rho$ and $\rho^{sep}$ share the same physical characters since
Bloch vector $\{r_i\}$ is the reflection of the intrinsic symmetry
in the systems \cite{he81}. While for $\cos\theta=-1$ one has
minimal overlap, and $\rho^{sep}$ shows distinct properties from
that of $\rho$.

In contrast to the Bures fidelity\cite{bures}, the overlap $\Lambda$
have clear geometric meaning whether for pure or mixed state, as
shown above. Furthermore by this geometric meaning, the optimal
procession can be reduced to find the fully separable state
$\rho^{sep}$ sharing the same physical properties with $\rho$ (see
Appendix A for a proof). Moreover this definition is more popular
than Eq.\eqref{purelap}. First $\Lambda$ comes back to the form
Eq.\eqref{purelap} for pure states. Second Eq.\eqref{overlap}
includes the case when one state is pure and the other is mixed.
This situation always happens as exemplified in Ref.\cite{cw07}, but
is not covered in the original discussion\cite{wg03}. Thirdly for
mixed state the geometric characters of GE becomes ambiguous because
of the \emph{convex roof} construction \cite{wg03}, while the
geometric meaning of $\Lambda$ is clear whenever for pure or mixed
state.

With these advantages, the evaluation of $\Lambda$ however is
difficult for mixed state $\rho^{sep}$ since there are infinite
possibilities of the decomposition for $\rho^{sep}$. Recently we
note a popular concept in condensed matter physics-Anderson's
Orthogonality Catastrophe(AOC)\cite{anderson67, mahan}, which refers
to the vanishing of the overlap between the many-body ground states
with and without the potential as a power law in the number of
particles in the systems. AOC is defined as
\begin{equation}\label{aoc}
\Delta=|\inp{\Phi}{\Phi^p}|^2,
\end{equation}
where $\ket{\Phi^p}$ and $\ket{\Phi}$ correspond respectively to the
many-body states with the potential and the state described entirely
in terms of free plane waves, including the ground state of the
unperturbed system\cite{anderson67}. Anderson proved that the
overlap $\Delta$ approached to be zero under thermodynamic limit
$N\rightarrow\infty$ even for very weak potential, that means that
$\ket{\Phi}$ is orthogonal to $\ket{\Phi^p}$\cite{anderson67} and
the transition between the two different states is forbidden. It is
the physical meaning of \emph{catastrophe}\cite{mahan}. As claimed
in Ref.\cite{anderson67}, it becomes impossible because of the
appearance of catastrophe to find the characters for many-body
systems by adiabatically imposing the potential and observing the
response, since the significant changes in many-body systems can be
induced even for infinitesimal perturbation. AOC presents an
understanding of a number of Fermi-edge singularities, e.g., in the
Kondo effect\cite{hewson} or in the X-ray edge problem\cite{aoc},
for which the local singularity has an overall effect on the
property of the many-body systems. With these points AOC manifestly
shows that the correlation in many-body systems can be constructed
simultaneously whenever the interaction appears, and thus can be
used to give a full description of correlation in many-body systems.
Furthermore the important feature is that this prohibition can be
conquered by the symmetry-breaking process as exemplified by the
observation of the X-ray absorption in the electron gas\cite{mahan},
which means that AOC can also be used to characterize the phase
transitions induced by the symmetry breaking process. In a word AOC
presents an comprehensive description of the many-body effects.

This crucial observation enforces us to define the following fully
separable state for $N$ parties,
\begin{equation}\label{rhos}
\rho^s=\rho_1\otimes\rho_2\otimes\cdots\otimes\rho_N,
\end{equation}
which represents the many-body state without potential, compared to
the state $\ket{\Phi}$ in Eq.\eqref{aoc}. And then we can define the
overlap with fully separable state $\rho^{s}$ to capture the overall
correlation in many-body systems,
\begin{equation}\label{delta}
\Delta=\max_{\{\rho^s\}}\text{Tr}[\rho\rho^s].
\end{equation}
This definition is the main contribution in this paper, and has some
distinct advantages, summarized as following
\begin{enumerate}
\item $\Delta$ has clear geometric meaning, which depicts the
minimal overlap angle between the Bloch vectors $\{r_i\}$ and
$\{r_i\}_s$. And for pure states, it returns to the original
definition Eq.\eqref{purelap} of GE.
\item By this geometric meaning, the optimal process in $\Delta$ can be reduced
to find the fully separable state $\rho^s$ sharing the same physical
features with $\rho$.
\item $\Delta$ can be regarded as the generalization of AOC to mixed state, and can
faithfully reflect the overall correlation in many-body systems.
\end{enumerate}

It should emphasize that this definition does not try to present a
complete measurement of the multipartite entanglement. Instead our
purpose is to find an universal method to characterize the overall
correlation in many-body systems, whether quantum or classical.
However this definition is also meaningful to find the unified
understanding of multipartite entanglement in many-body systems.  As
shown in the next section, $\Delta$ indeed presents the interesting
information for the phase transition in Dicke model. And moreover
the connection between $\Delta$ and multipartite entanglement in
Dicke model has also been discussed in Sec.III. Additionally in
contrast to the recent interest in the fidelity for many-body
systems\cite{qz06}, $\Delta$ does not serve for the state
discrimination.

\section{Exemplification: Phase Transition in the Dicke Model}
In order to demonstrate the generality of this definition, the phase
transition in Dicke model is discussed by $\Delta$ in this section.
Dicke model describes the dynamics of $N$ independently identical
two-level atoms coupling to the same quantized electromagnetic
field\cite{dicke54}. Due to the presence of dipole-dipole force
between atoms, Dicke model shows the normal-superradiant
transition\cite{hh73}.

Dicke model is related to many fundamental issues in quantum optics,
quantum mechanics and condensed matter physics, such as the coherent
spontaneous radiation\cite{hh73}, the dissipation of quantum system
\cite{leggett87}, quantum chaos\cite{hakee} and atomic
self-organization in a cavity\cite{bgbe09}. The normal-superradiant
transition have been first observed with Rydberg atoms
\cite{raimond82}, and recently in a superfluid gas coupled to an
optical cavity\cite{bgbe09} and nuclear spin ensemble surrounding a
single photon emitter\cite{kylcg10}. Quantum entanglement in Dicke
model has also been discussed extensively in\cite{leb04, vdb07}.
Furthermore Dicke model is also related to the issues of how the
opened multipartite systems is affected by the environment and the
robustness of multipartite entanglement\cite{ctp09}

The Hamiltonian for single-model Dicke model reads
\begin{eqnarray}\label{dicke}
H&=&\omega a^{\dagger}a + \frac{\omega_0}{2}\sum_{i=1}^N\sigma_i^z +
\frac{\lambda}{\sqrt{N}}\sum_{i=1}^N
(\sigma^+_i+\sigma^-_i)(a^{\dagger}+a)\nonumber\\ &=& \omega_0 J_z +
\omega
a^{\dagger}a+\frac{\lambda}{\sqrt{N}}(a^{\dagger}+a)(J_++J_-),
\end{eqnarray}
where $J_z=\sum_{i=1}^N\sigma_i^z/2$ and
$J_{\pm}=\sum_{i=1}^N\sigma_i^{\pm}$ are the collective angular
momentum operators. At zero temperature, the normal-superradiant
transition happens when $\lambda=\lambda_c=\sqrt{\omega\omega_0}/2$.
For finite temperature, the critical temperature is decided by the
relation\cite{lz05}
\begin{equation}\label{tc}
\beta_c=\frac{\omega_0}{2\lambda^2}\frac{\tanh(\beta_c\omega/2)}{\tanh(\beta_c\omega_0/2)}.
\end{equation}
An intrinsic property of Dicke model is the parity symmetry,
\begin{eqnarray}\label{ps}
[H, \Pi]&=&0,\nonumber\\
\Pi&=&e^{i\pi(a^{\dagger}a+J_z+\frac{N}{2})}.
\end{eqnarray}
Moreover Eq.\eqref{dicke} is obviously permutation invariant by
exchanging any two atoms.

With these information, the overlap $\Delta$ for the Dicke model is
studied explicitly for zero and finite temperature in the following
two subsections. Some interesting features of $\Delta$ are
displayed.

\subsection{Zero Temperature}
With respect of the permutation invariance of atoms in Eq.
\eqref{dicke}, it is convenient to introduce the
Holstein-Primakoff(HP) transformation
\begin{eqnarray}\label{hp}
J_z&=&b^{\dagger}b-\frac{N}{2}\nonumber\\
J_+&=&b^{\dagger}\sqrt{N-b^{\dagger}b}\nonumber\\
J_-&=&\sqrt{N-b^{\dagger}b}b.
\end{eqnarray}
with bosonic operator $b^{(\dagger)}$. Semiclassically there is a
ground state with $J_z=-N/2$ for Dicke model under thermodynamic
limit $N\rightarrow\infty$. Hence it is reasonable to adopt the
low-excitation approximation at zero temperature, and then one
obtains two effective Hamiltonians for different regions of
$\lambda$ (refer to \cite{eb03} for details)
\begin{widetext}
\begin{eqnarray}\label{edicke}
H^{(1)}&=&\omega a^{\dagger}a+\omega_0
b^{\dagger}b+\lambda(a^{\dagger}+a)(b^{\dagger}+b)-\frac{N}{2}\omega_0, \hspace{2em} \lambda<\lambda_c;\nonumber\\
H^{(2)}&=&\omega a^{\dagger}a+[\omega_0 +
\frac{2}{\omega}(\lambda^2-\lambda_c^2)]b^{\dagger}b+\frac{(\lambda^2-\lambda_c^2)(3\lambda^2+\lambda_c^2)}{2\omega(\lambda^2+\lambda_c^2)}(b+b^{\dagger})^2
\nonumber\\&&+\frac{\sqrt{2}\lambda_c^2}{\sqrt{\lambda^2+\lambda_c^2}}(a^{\dagger}+a)(b^{\dagger}+b)
+\text{const.},  \hspace{4em}\lambda>\lambda_c.
\end{eqnarray}
\end{widetext}
$H^{(1)}$ and $H^{(2)}$ can be diagonalized readily by transforming
into phase space, and then one has the diagonalized form \cite{eb03}
\begin{equation}
H=\omega_1 c_1^{\dagger}c_1 + \omega_2c_2^{\dagger}c_2.
\end{equation}
where the forms of $\omega_{1(2)}$ and $c_{1(2)}$ are dependent on
$\lambda>\lambda_c$ or $\lambda<\lambda_c$\cite{eb03}. Then the
ground state can be written as
\begin{equation}\label{gs}
\ket{g}=\ket{g}_1\otimes\ket{g}_2,
\end{equation}
where $\ket{g}_{1(2)}$ denotes the vacuum state for mode
$\omega_{1(2)}$. Furthermore the average spin along $z$ direction
per atom shows distinct values across the phase transition point,
\begin{equation}\label{jz}
\frac{\langle J_z\rangle}{N}=
\begin{cases}-\frac{1}{2},  & \lambda<\lambda_c;\\
-\frac{\lambda^2_c}{2\lambda^2}, & \lambda>\lambda_c,
\end{cases}
\end{equation}
which then can act the order parameter. It is obvious that a
macroscopic number of atoms are excited for $\lambda>\lambda_c$,
which is so called superradiant phase, while for $\lambda<\lambda_c$
it is normal phase.

With these information, we are ready to evaluate $\Delta$. Our focus
is mainly on the atom system. Then the crucial step is to decide the
fully separable state $\rho^s$ for atom system. As mentioned in
Sec.II and proved in Appendix A, the optimum process in Eq.\eqref{D}
can be reduced to find $\rho^s$ sharing the same global features
with the ground state Eq.\eqref{gs}. First with the requirement of
the permutation invariance of atoms in Dicke model, the single
atomic state should have the same form in $\rho^s$, i.e.
$\rho_i=\varrho, i=1,2,\cdots, N$, and then
\begin{equation}
\rho^s=\varrho^{\otimes N}.
\end{equation}
Second the parity symmetry for Dicke model is reduced for single
atom state $\rho$ as
\begin{eqnarray}
[e^{i\pi J_z}, \rho^s]=0\nonumber\\
\Rightarrow [e^{i\pi\sigma_z}, \varrho]=0.
\end{eqnarray}
Thus one has under $\sigma_z$ representation
\begin{equation}\label{srho}
\varrho=\left(\begin{array}{cc}a&0\\0&1-a\end{array}\right).
\end{equation}
Finally with requirement of Eq.\eqref{jz},  $a=1/2+\frac{\langle
J_z\rangle}{N}$. Thus $\rho^s$ can be uniquely determined as
\begin{equation}
\rho^s=\left(\begin{array}{cc}1/2+\frac{\langle
J_z\rangle}{N}&0\\0&1/2-\frac{\langle
J_z\rangle}{N}\end{array}\right)^{\otimes N}.
\end{equation}
It should point out that the procedure for the determination of
$\rho^s$ is popular whether for zero or finite temperature.

\begin{figure}[t]
\center
\includegraphics[bbllx=17, bblly=19, bburx=275, bbury=216, width=8cm]{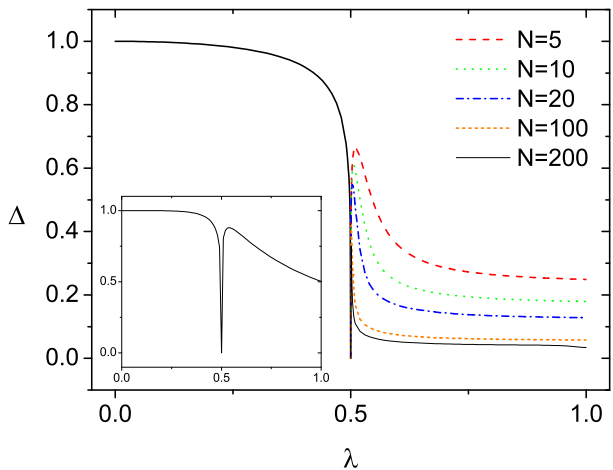}
\caption{\label{0d}The overlap $\Delta$ with fully separable state
$\rho^s$ vs. the coupling $\lambda$ at zero temperature.
$\omega_0=\omega=1$ has been chosen for this plotting, and the
critical point is $\lambda_c=0.5$ in this case. The inset is a
plotting for the purity of the reduced density of atomic freedom
under $N\rightarrow\infty$.}
\end{figure}

For the evaluation of the overlap $\Delta$, it should note that
$\rho^s$ can be rewritten as the following contract form under the
$J_z$ representation,
\begin{equation}
\rho^s=\sum_{n=1}^{N}{_N C_n}a^n
(1-a)^{N-n}\ket{n-\frac{N}{2}}\bra{n-\frac{N}{2}},
\end{equation}
where ${_N C_k}$ denotes the binomial function, and
$\ket{n-\frac{N}{2}}$ presents the state for which $n$ particles is
spin-up and the other is spin-down. Together with the HP
transformation Eq.\eqref{hp}, it is obvious
\begin{equation}
b^{\dagger}b\ket{n-\frac{N}{2}}=n\ket{n-\frac{N}{2}}.
\end{equation}
Then  $\Delta$ can be evaluated easily under this representation.

In Fig.\ref{0d}, the overlap $\Delta$ with $\rho^s$ is plotted. At
normal phase ($\lambda<\lambda_c$), one has $\langle
J_z\rangle/N=-0.5$, $a=0$, and then $\rho^s$ is the fully separable
pure state.  $\Delta$ is determined mainly by the first diagonal
element of the reduced density matrix of atom system in this special
case. While for $\lambda>\lambda_c$, $\Delta$ shows a sudden rising
and then decreases with the increment of $\lambda$, and tends to be
steady with $\lambda\rightarrow \infty$. Moreover under
$N\rightarrow\infty$, $\Delta$ tends to be vanishing.  Then two
different phases can be clearly identified by evaluating $\Delta$.

Some intricate features of the phase transition can be disclosed by
$\Delta$. For normal phase $\lambda<\lambda_c$, it is known that the
atom system becomes entangled with the  electromagnetic field, and
attains the maximal value at the critical point\cite{leb04}. The
entanglement leads the state of atom system to be mixed, and the
purity of its reduced density is decreased as shown by the inset in
Fig.\ref{0d}. At the same time the pairwise entanglement between two
any atoms is also raised mediated by their couplings to the
electromagnetic field, and the atoms become correlated with each
other \cite{leb04}. These intrinsic properties can  be captured by
$\Delta$ at the same time. For normal phase $\rho^s$ is pure and
fully separable. Thus the decrement of $\Delta$ reflects the fact
that the atoms become correlated with each other, and attains the
maximal correlation at the critical point, at which $\Delta$ has
minimal value. Furthermore since there is no interaction among
atoms, the only reason for the construction of correlation in atoms
is the couplings to the same electromagnetic field, which just
induces the state for atom system to be mixed. This feature can also
manifested by the decrement of $\Delta$ with respect that $\rho^s$
is pure.

For superradiant phase $\lambda>\lambda_c$, it is known that the
entanglement between the atoms and electromagnetic field decreases
monotonously to a steady value with the increment of $\lambda$,
while the pairwise entanglement in atoms disappears
asymptotically\cite{leb04}. Contrastably the purity for the state of
atom system has a sudden increasing closed to $\lambda_c$ and then
decreases to a steady value, as shown by the inset of Fig.\ref{0d}.
The two different behaviors can also be captured by $\Delta$.
Similar to the behavior of the purity of the state for atom system,
$\Delta$ has also a sudden arising closed to $\lambda_c$ and then
decreases to a steady value with the increment of $\lambda$. With
respect that $\rho^s$ is mixed in this case and its purity is
monotonically decreased with the increment of $\lambda$, the abrupt
increment of $\Delta$ means that the sudden recovery of the purity
of atomic system is at the expense of the reduction of correlation
in atoms. It is obvious from Fig.\ref{0d} that $\Delta$ tends to be
zero with the increment of $N$ for large $\lambda$. However the
vanishing of $\Delta$ cannot attribute to the mixedness of $\rho^s$
since the steady value of $\Delta$ for finite $N$ is always bigger
than the maximal mixedness $1/N$, manifested by Fig.\ref{0d}. This
feature means that the correlation in atoms still exists. Since the
pairwise entanglement of atoms is known to be vanished in this
limit\cite{leb04}, the correlation in atoms must be global.

The scaling behavior of $\Delta$ near the critical point show some
interesting features. At the normal phase ($\lambda<\lambda_c$), one
has for $\omega=\omega_0=1$
\begin{equation}
\Delta=\frac{2^{3/2}(1-4\lambda^2)^{1/4}}{[1+3\sqrt{1-4\lambda^2}+0.5(\sqrt{1+2\lambda}+\sqrt{1-2\lambda})^3]}.
\end{equation}
Similar to the method in Ref.\cite{orus08}, one can define the globe
overlap $-\ln\Delta$ to measure the atomic correlation in Dicke
model. It is obvious that the globe overlap is mainly determined by
$(1-4\lambda^2)^{1/4}$ near $\lambda_c=1/2$, and then
\begin{equation}
-\ln\Delta\sim-\tfrac{1}{4}\ln(1-\tfrac{\lambda}{\lambda_c}),
\end{equation}
which is same to the scaling behavior of multipartite entanglement
in the Lipkin-Meshkov-Glick(LMG) model\cite{orus08}. This result is
not strange since Dicke model and LMG model belong to the same
universality class. However it strongly implies that $\Delta$ could
be correlated directly to the multipartite entanglement in Dicke
model. As shown in Sec.IV, the atom system indeed displays the
multipartite entanglement in this case.

\subsection{Finite Temperature}
\begin{figure}[tbp]
\center
\includegraphics[bbllx=15, bblly=1, bburx=307, bbury=187,
width=7cm]{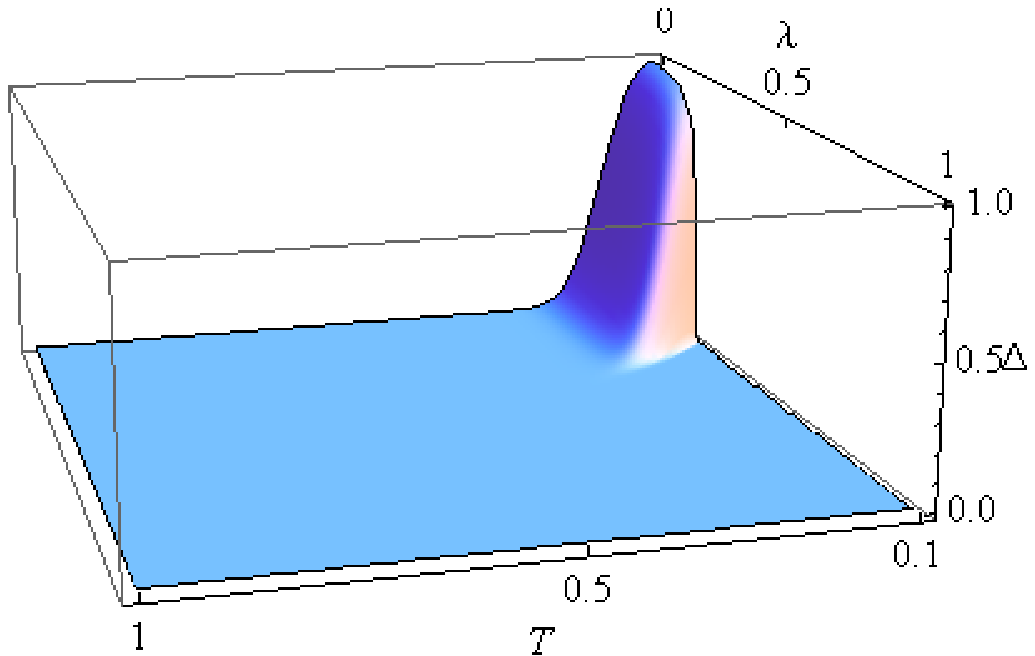}\vspace{1.5em}
\includegraphics[bbllx=15, bblly=1, bburx=307, bbury=187, width=7cm]{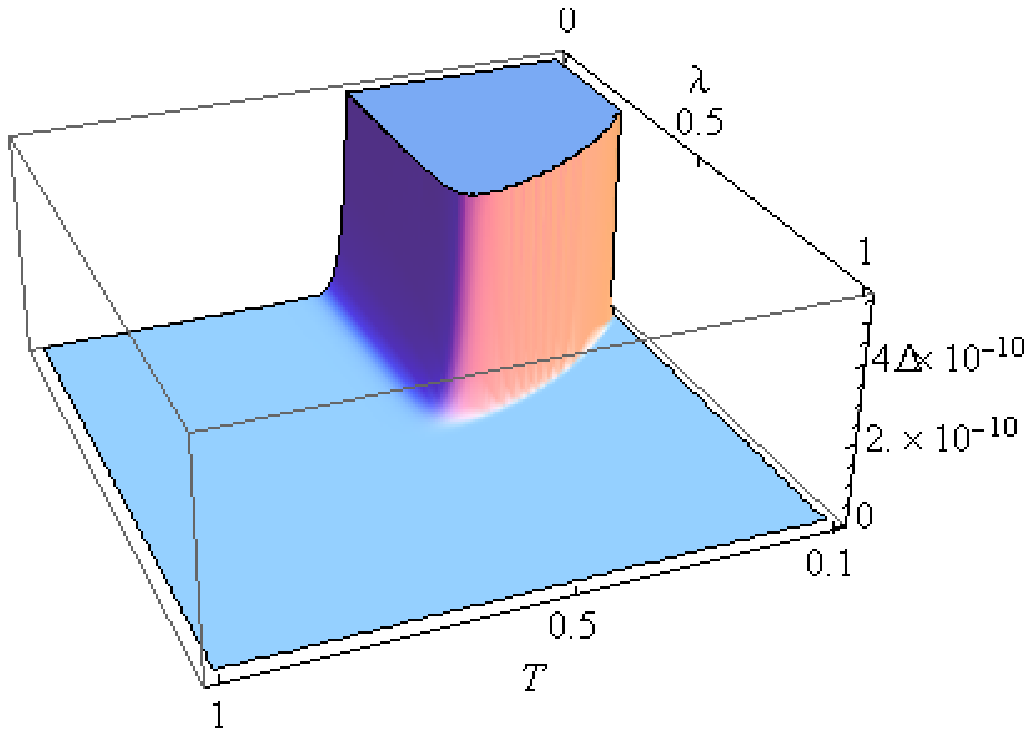}
\vspace{1.5em}
\includegraphics[bbllx=2, bblly=3, bburx=299, bbury=299, width=6cm]{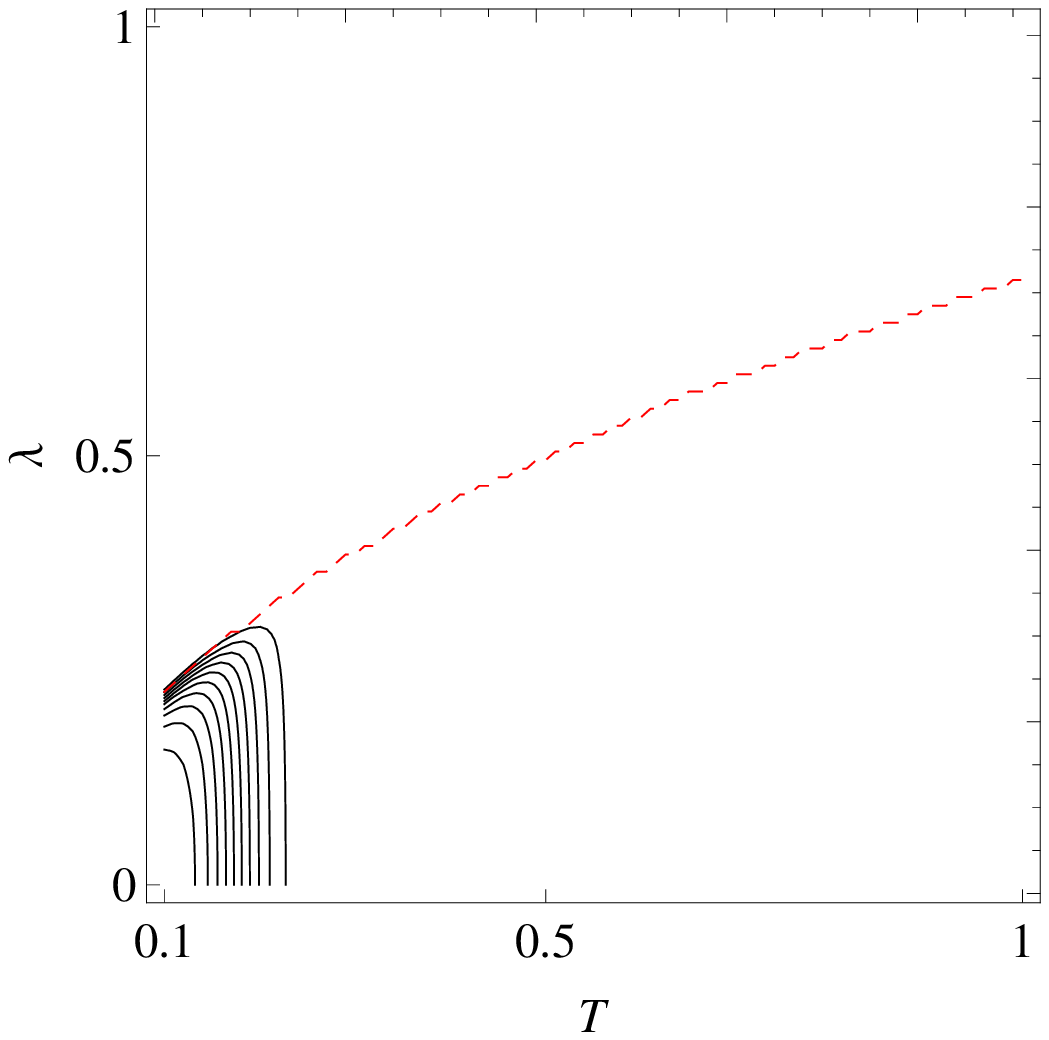}
\caption{\label{td}Overlap $\Delta$ with fully separable state
$\rho^s$ vs. the coupling $\lambda$ and temperature $T$.
$\omega_0=\omega=k_B=1$ and $N=100$ has been chosen for this
plotting. The two 3-dimensional figures are the same plotting with
different plot-ranges for clarity. In the contour plotting the red
dashed line corresponds to the critical line
$T_c=2\lambda^2/k_B\omega_0$, and because of the rapid decay of
$\Delta$ only finite range of its values is shown for this contour
plot.}
\end{figure}

At finite temperature the phase transition is induced by thermal
fluctuation. In order to determine the critical temperature, the
general method is to evaluate the partition function $z$. In
Ref.\cite{lz05}, $z$ has been obtained analytically
\begin{eqnarray}
z&=&\frac{\sqrt{1/2\pi}}{1-e^{-\beta\omega}}\int_{-\infty}^{\infty}\text{d}x
e^{-\frac{x^2}{2}}\nonumber\\ &\times&
\left\{2\cosh[\beta\sqrt{\frac{\omega_0^2}{4}+\frac{x^2\lambda^2}{N}\coth\frac{\beta\omega}{2}}]\right\}^N
\end{eqnarray}
and the critical temperature is determined by Eq.\eqref{tc}. For
$\omega=\omega_0$, it is reduced to $T_c=2\lambda^2/k_B\omega_0$.
With the same trick used in \cite{lz05}, the overlap $\Delta$ can
also be written analytically as (see Appendix B for the details of
calculation)
\begin{eqnarray}\label{toverlap}
\Delta=\frac{1}{z}\frac{\sqrt{1/2\pi}}{1-e^{-\beta\omega}}\int_{-\infty}^{\infty}\text{d}x
e^{-\frac{x^2}{2}}&\nonumber\\
\times\left\{2\cosh\left[\beta\sqrt{\frac{\omega_0^2}{4}+\frac{x^2\lambda^2}{N}\coth\frac{\beta\omega}{2}}\right]
\right.+&\nonumber\\
\left.\tfrac{\omega_0(1-2a)/2}{\sqrt{\tfrac{\omega_0^2}{4}+\tfrac{x^2\lambda^2}{N}\coth\tfrac{\beta\omega}{2}}}
\sinh\left[\beta\sqrt{\tfrac{\omega_0^2}{4}+\tfrac{x^2\lambda^2}{N}\coth\tfrac{\beta\omega}{2}}\right]\right\}^N&
\end{eqnarray}

As shown in Fig.\ref{td}, $\Delta$ can clearly detect the phase
transition by its abrupt variance closed to the critical line. With
respect that $\rho^s$ is mixed and fully separable in this case,
$\Delta$ reflects that the correlation in atoms exists even for
finite temperature. However this type of correlation is obviously
induced by the thermal fluctuation, and thus is incoherent in
contrast to that for zero temperature. This difference will become
clear if one focuses on the multipartite entanglement of atoms in
the next section.

\section{Multipartite Entanglement in Dicke Model}
Another interesting aspect for Dicke model is the multipartite
entanglement in atoms. Since all atoms simultaneously couple
isotropically to the same electromagnetic field, then it is expected
that the multipartite correlation of atoms could be readily
constructed in this case.

However the measure of multipartite entanglement is a difficult task
in general, especially for mixed state. An indirect way of resolving
this difficulty is to find the characters uniquely belonging to the
fully separable state Eq.\eqref{sep}, and the violation of these
properties implies the appearance of multipartite entanglement. Spin
squeezing is one of the most successful approaches to the
multipartite entanglement in this way\cite{sdcz01}. Recently G.
T\'oth, {\it et.al.} provides a series of inequalities about spin
squeezing to identify the multipartite entanglement in collective
models\cite{tkgb09},
\begin{subequations}
\begin{eqnarray}
\exs{J_x^2}+\exs{J_y^2}+\exs{J_z^2} &\le& \tfrac{N(N+2)}{4},
\\
\Delta^2J_x+\Delta^2J_y+\Delta^2J_z &\ge& \tfrac{N}{2},
\\
\exs{J_{\alpha}^2}+\exs{J_{\beta}^2}-\tfrac{N}{2} &\le&
(N-1)\Delta^2J_{\gamma},
\\
(N-1)\left[\Delta^2J_{\alpha}+\Delta^2J_{\beta}\right] &\ge&
\exs{J_{\gamma}^2}+\tfrac{N(N-2)}{4},
\end{eqnarray}
\end{subequations}
where $\alpha, \beta, \gamma$ adopt the all permutation of $x, y,
z$, and $\Delta^2J_{\alpha}=\langle J_{\alpha}^2 \rangle-\langle
J_{\alpha} \rangle^2$. The violation of any one of these
inequalities implies the appearance of entanglement\cite{tkgb09}.
With respect of the limit large $N$, these inequalities can be
rewritten as
\begin{subequations}\label{ineq}
\begin{eqnarray}
\tfrac{1}{N^2}(\exs{J_x^2}+\exs{J_y^2}+\exs{J_z^2} )&\le&
\tfrac{1}{4}, \label{ineqa}
\\
\tfrac{1}{N^2}(\Delta^2J_x+\Delta^2J_y+\Delta^2J_z)-\tfrac{1}{2N}
&\ge& 0, \label{ineqb}
\\
\tfrac{\Delta^2J_{\gamma}}{N}-
\tfrac{1}{N^2}(\exs{J_{\alpha}^2}+\exs{J_{\beta}^2})+\tfrac{1}{2N}
&\ge& 0, \label{ineqc}
\\
\tfrac{1}{N}(\Delta^2J_{\alpha}+\Delta^2J_{\beta} )-
\tfrac{\exs{J_{\gamma}^2}}{N^2}-\tfrac{1}{4}&\ge& 0, \label{ineqd}
\end{eqnarray}
\end{subequations}
in which $\tfrac{1}{N^2}\exs{J_{\alpha}^2}$ and
$\tfrac{1}{N^2}\Delta^2J_{\alpha}$ are equivalent to evaluate the
average $\exs{(J_{\alpha}/N)^2}$ and
$\Delta^2(J_{\alpha}/N)=\exs{(J_{\alpha}/N)^2}-\exs{J_{\alpha}/N}^2$.
For large $N$, these inequalities have nontrivial result since the
average magnetization per particle and its fluctuation still have
nonvanishing values. It should point that Eq.\eqref{ineqa} is
obviously satisfied for arbitrary state. So the following discussion
is mainly about Eqs.\eqref{ineq}(b-d).

\subsection{Zero Temperature }

\begin{figure}[t]
\center
\includegraphics[bbllx=17, bblly=19, bburx=275, bbury=216, width=7cm]{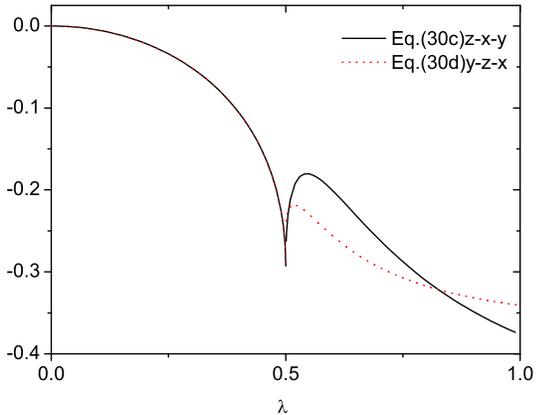}
\caption{\label{ineq0} Eqs.\eqref{ineq} vs. the coupling $\lambda$
at zero temperature. $\omega_0=\omega=1$ and $N=100$ have been
chosen for this plotting. The labels $z-x-y$ and $y-z-x$ denote the
sequence and values of $\alpha-\beta-\gamma$ in corresponding
inequalities. }
\end{figure}

The evaluations of $\exs{J_{\alpha}/N}$ and $\exs{(J_{\alpha}/N)^2}$
can be implemented readily through Bogoliubov
transformation\cite{eb03}. Our calculations show that
Eq.\eqref{ineqb} is always satisfied at both normal and superradiant
phases. In Fig.\ref{ineq0}, several situations for Eqs.\eqref{ineq}
have been plotted with limit $N\rightarrow\infty$, and the others
can be proved to be bigger than zero. The violation implies that the
atoms should be entangled. Moreover since the pairwise entanglement
between atoms is known to be vanished with increment of
$\lambda$\cite{leb04}, this entanglement is sure to be multipartite.
Furthermore there is also a sudden increment closed to the critical
point, similar to the behavior of $\Delta$ shown in Fig.\ref{0d}.
This feature means that there is a sudden reduction of the
correlation of atoms, and $\Delta$ can also be used to detect the
entanglement of atoms in Dicke model at zero temperature.

\subsection{Finite Temperature}
\begin{figure}[tbp]
\center
\includegraphics[bbllx=24, bblly=1, bburx=260, bbury=187,width=4.5cm]{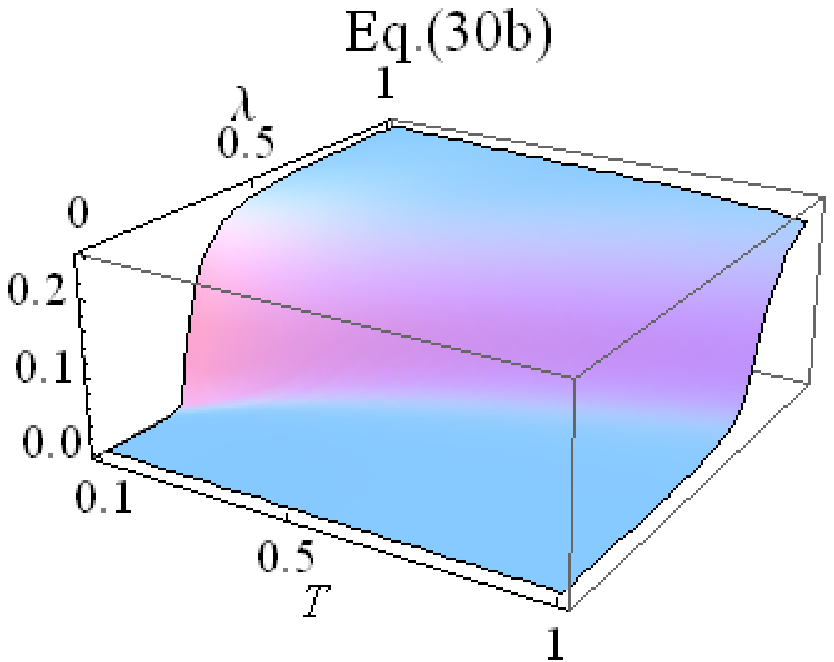}
\hspace{1em}\vspace{1em}
\includegraphics[bbllx=22, bblly=1, bburx=272, bbury=172,width=4.1cm]{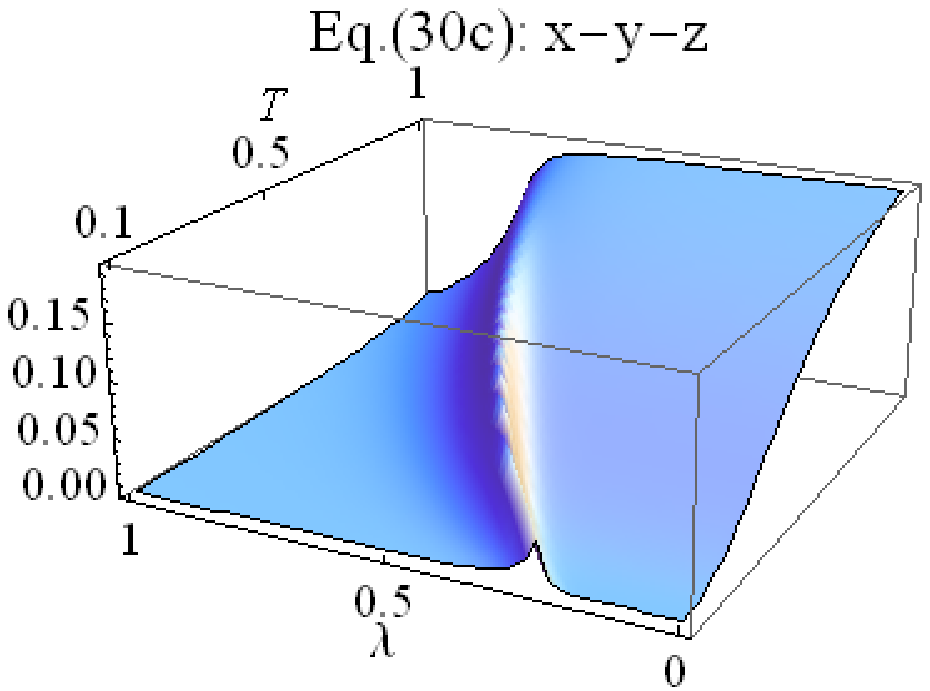}
\hspace{1em}\vspace{1em}
\includegraphics[bbllx=8, bblly=0, bburx=278, bbury=203,width=4.1cm]{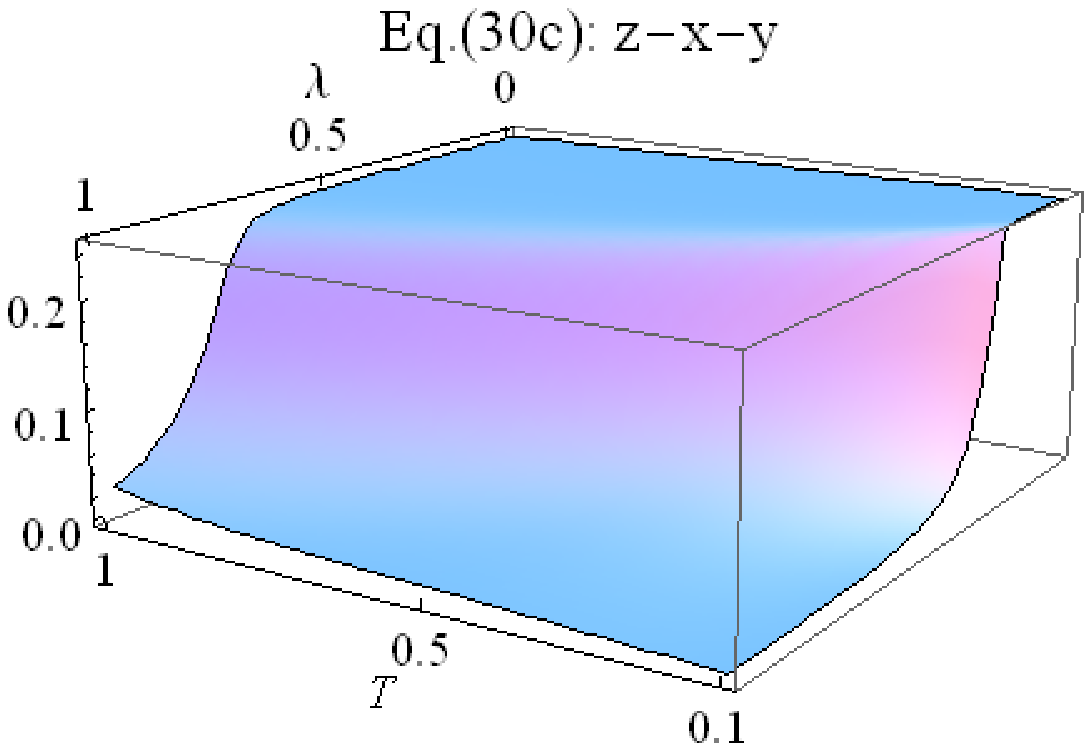}
\hspace{1em}\vspace{1em}
\includegraphics[bbllx=11, bblly=1, bburx=273, bbury=219,width=4.1cm]{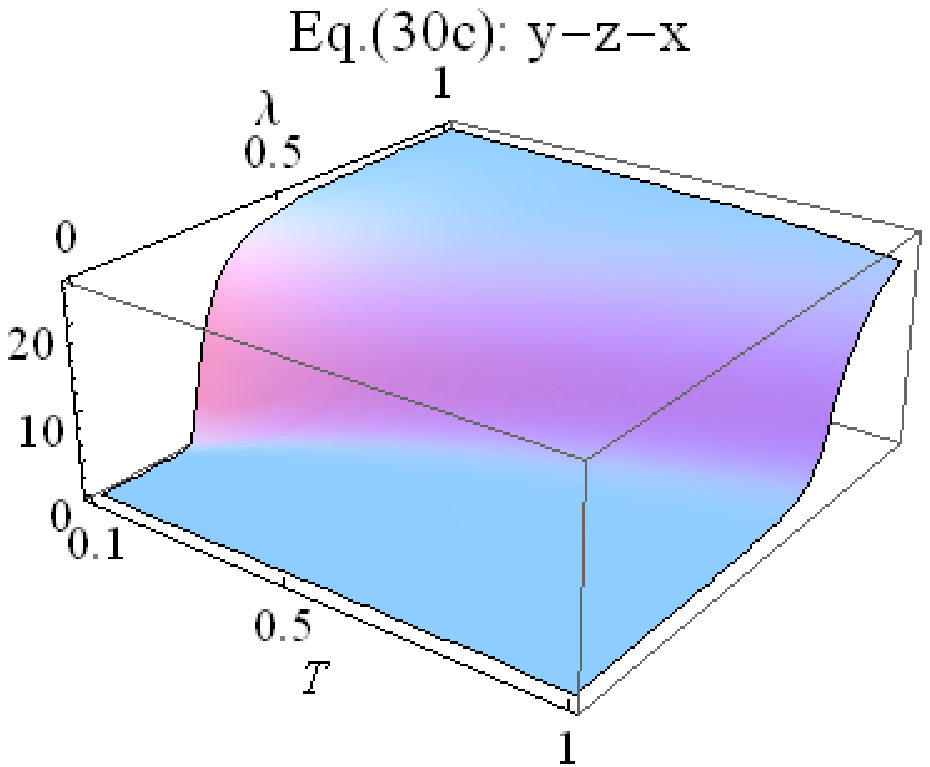}
\hspace{1em}\vspace{1em}
\includegraphics[bbllx=0, bblly=0, bburx=288, bbury=232,width=4.1cm]{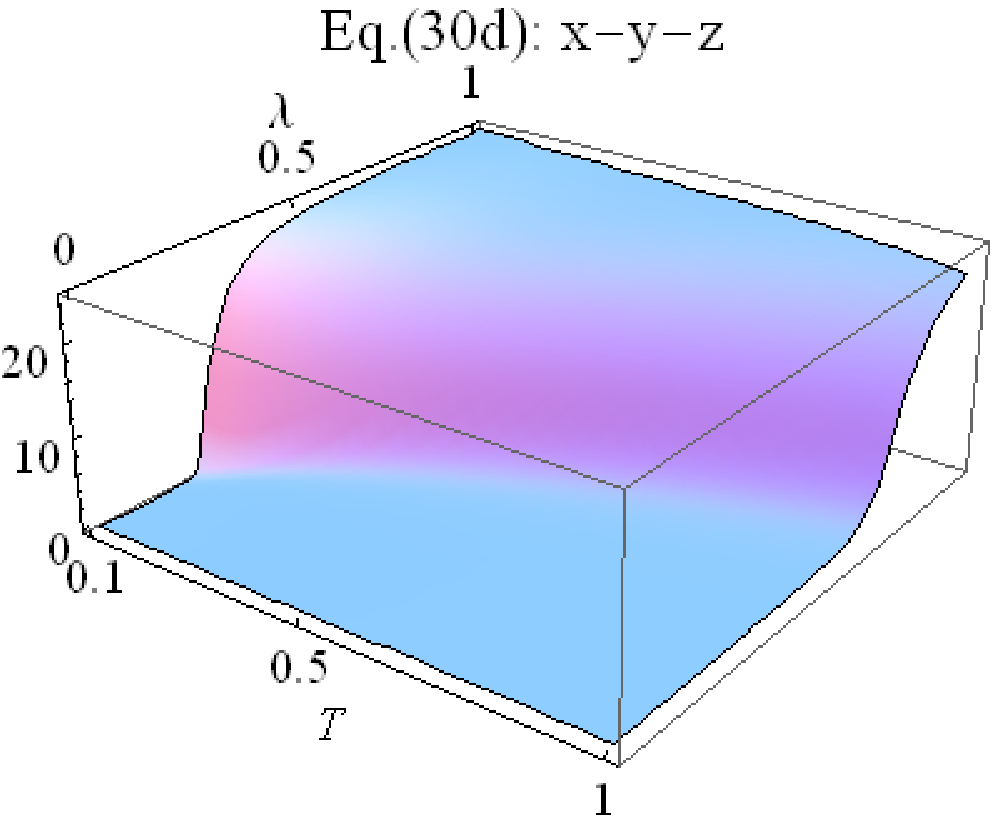}
\hspace{1em}\vspace{1em}
\includegraphics[bbllx=9, bblly=1, bburx=277, bbury=220,width=4.1cm]{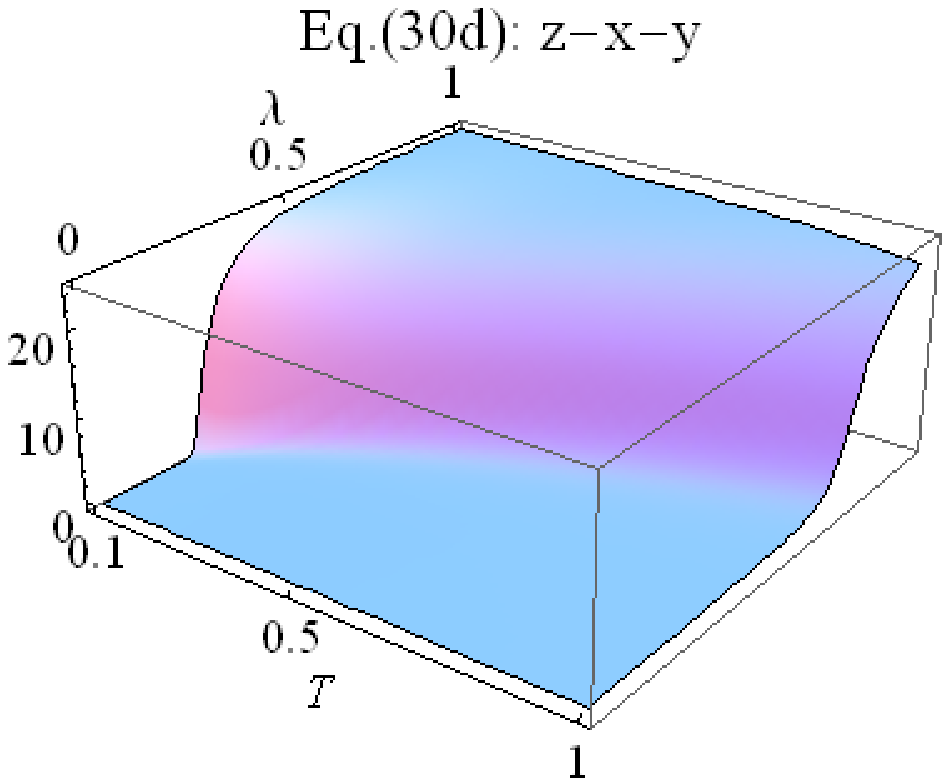}
\hspace{1em}\vspace{1em}
\includegraphics[bbllx=16, bblly=0, bburx=266, bbury=181,width=4.1cm]{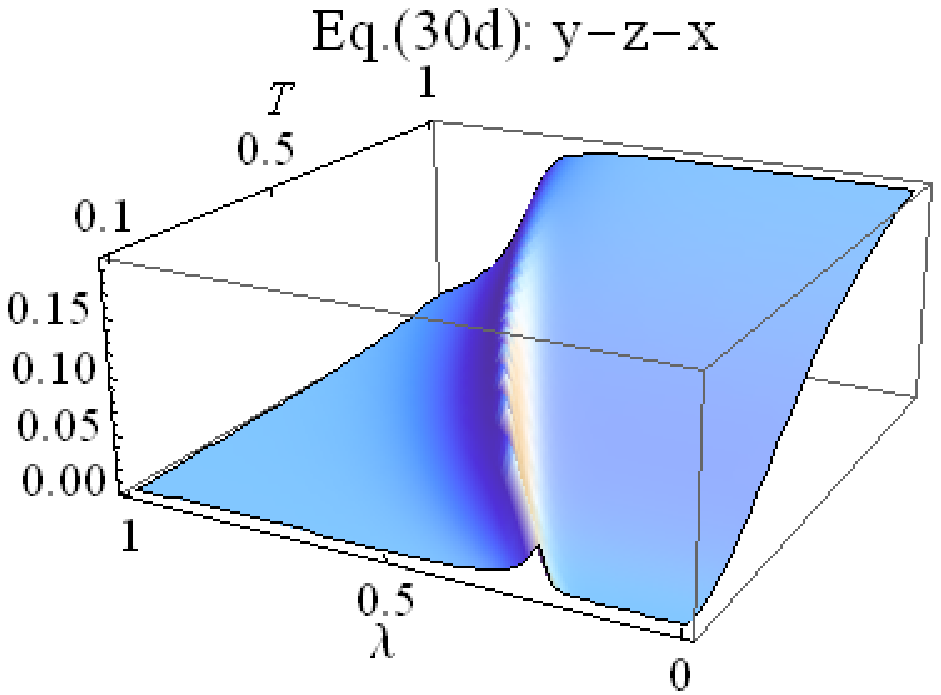}
\caption{\label{ineqT} Eqs.\eqref{ineq} vs. the coupling $\lambda$
and temperature $T$. $\omega_0=\omega=k_B=1$ and $N=100$ has been
chosen for these plots. The labels x-y-z$, $$z-x-y$ and $y-z-x$
denote the sequence and values of $\alpha-\beta-\gamma$ in
corresponding inequalities. The similarity among several plots is
because their corresponding inequalities respectively would become
closed under limit $N\rightarrow\infty$. }
\end{figure}

At finite temperature, the evaluations of $\exs{J_{\alpha}/N}$ and
$\exs{(J_{\alpha}/N)^2}$ can adopt the same trick used in
Ref.\cite{lz05} (also shown in Appendix B). In Figs.\ref{ineqT},
Eqs.\eqref{ineq}(b-d) have been plotted with all possible
permutation of $x, y, z$. It is obvious that all inequalities are
satisfied simultaneously, and then one can conclude that there is no
quantum entanglement of atoms in this case. This result is not
surprising since the thermal fluctuation is dominant at finite
temperature, and is considered to be incoherent. Although the
absence of quantum correlation, the correlation induce by thermal
fluctuation predominates, as shown in Figs.\ref{td} by $\Delta$,
which means that $\Delta$ can also be used to detect the thermal
correlation.

\section{Conclusions and Further Discussion}
In this paper, the overlap $\Delta$ with a special fully separable
state defined in Eq.\eqref{delta} is introduced, in order to capture
the overall correlation in many-body systems, whether quantum or
classical. $\Delta$ has clear geometric and physical meaning shown
in Sec.II. With these features, the optimum process in the
definition of $\Delta$ can be reduced to find the fully separable
state $\rho^s$ defined in Eq.\eqref{rhos}, which shares the same
physical properties with the state to be measured. Importantly
$\Delta$ can be considered as the generalization of the concept
Anderson's Orthogonality Catastrophe\cite{anderson67, mahan}, which
is critical for the understanding of some effects in condensed
matter physics. This important connection displays the popularity of
$\Delta$ to detect the global correlation in many-body systems. And
as an exemplification the phase transition in Dicke model has been
discussed by $\Delta$.

As shown in Sec.III, $\Delta$ unambiguously depicts the phase
transition features and the global correlation in Dicke model,
whether for zero or finite temperature. At zero temperature,
$\Delta$ displays the distinct behaviors across the critical point.
Furthermore with the information of $\rho^s$, $\Delta$ predicts the
appearance of the multipartite entanglement in atom system, as
verified in Sec.IVA.

As for finite temperature, $\Delta$ can still be used to mark the
phase transition in Dicke model, as shown in Figs.\ref{td}. It
displays the sudden variance at the critical line decided by the
temperature $T$ and the coupling $\lambda$. An intricate feature
appears when $T\rightarrow\infty$. It is believed that all atoms
would becomes independent in this case and can be considered as the
ideal system\cite{pathria}. Quantum mechanically, the state in this
case can be described by a fully separable state. Whereas $\Delta$
approaches zero as shown in Figs.\eqref{td}, and the non-zero
$\Delta$ appears only at intermediate temperature, shown in
Figs.\eqref{td}. This phenomenon implies that the correlation in
atoms would exist even under high temperature. Moreover under the
$J_z$ representation, the dimension is proportional to the atomic
number $N$, and the value of the overlap shown in Figs. \eqref{td}
has exceeded greatly the limit by $N$. Thus this phenomenon cannot
attribute to the mixedness of the state for atom system.
Unfortunately we do not know how to understand these two different
features.

Although $\Delta$ cannot present a complete measurement of
multipartite entanglement, it has been shown the intimate connection
to the quantum entanglement in some special cases, such as the
discussion for Dicke model at zero temperature in this paper. From
this discussion $\Delta$ would presents a complete description for
the global correlation in many-body systems, whether quantum or
classical. So it is not surprising that $\Delta$ can be used to
identify the quantum entanglement in some special cases. However it
is difficult to answer the question what the general relation
between $\Delta$ and quantum entanglement is since the absence of
the unified understanding of multipartite entanglement. This point
will be studied in the future publication.

\subsection*{Appendix A: Find the nearest $\rho^{s}$ for a definite $\rho$ }
For two arbitrary density matrixes $\rho_1$ and $\rho_2$, they
always has the following decompositions simultaneously
\begin{eqnarray}
\rho_1&=&\sum_{n}p^{(1)}_n\ket{n}_1{_1\bra{n}}\nonumber\\
\rho_2&=&\sum_{m}p^{(2)}_m\ket{m}_2{_2\bra{m}},
\end{eqnarray}
where $p^{1(2)}_{n(m)}$ denotes the probability that the system is
being in the state $\ket{n(m)}_{1(2)}$. It should emphasize that it
is \emph{unnecessary} for the states labeled by different $n$ or $m$
to be orthogonal with each other. And thus the decompositions above
can always be realized at the same time. Then the overlap between
$\rho_1$ and $\rho_2$ reads
\begin{equation}
\text{Tr}[\rho_1\rho_2]=\sum_{m,n}p^{(1)}_n
p^{(2)}_m|{_2\inp{m}{n}_1}|^2.
\end{equation}
Obviously the maximization of overlap is dependent on the inner
product $|{_2\inp{m}{n}_1}|^2$. It is well known that for two
\emph{different} states $\ket{v}$ and $\ket{w}$ their inner product
is bounded by Cachy-Schwartz(CS) inequality, i.e.,
\begin{equation}
|\inp{v}{w}|^2\leq\inp{v}{v}\inp{w}{w},
\end{equation}
where the equality occurs if and only if the two vectors $\ket{v}$
and $\ket{w}$ in Hilbert state are linearly related, i.e.
$\ket{v}=c\ket{w}$ for some scalar $c$. The important point for this
condition is that $c$ is \emph{not} necessary a constant, for which
$\ket{v}$ and $\ket{w}$ become physically identical, and CS
inequality has trivial consequence. Thus
\begin{equation}
\text{Tr}[\rho_1\rho_2]\leq\sum_{m,n}p^{(1)}_n
p^{(2)}_m{_1\inp{n}{n}_1}{_2\inp{m}{m}_2},
\end{equation}
where the equality occurs if and if only for arbitrary $\ket{n}_1$
and $\ket{m}_2$ they are still linearly related. But in this case
the scalar $c$ has to be dependent on both $n$ and $m$, i.e.
$c=c_{mn}$, which means that any $\ket{n}_1$ have to be linearly
related to all $\ket{m}_2$. An interesting consequence for this
condition is $[\rho_1, \rho_2]=0$, which means that $\rho_1$ and
$\rho_2$ share the same set of eigenvectors, and thus they share the
same global symmetry and belong to the same space.

This conclusion is not strange if one notes that the overlap between
two matrixes is mainly determined by the inclusion relation of the
spaces decided by the matrixes. As an example let consider two
matrixes belong to two completely different spaces. And then the
overlap must be zero since mathematically the intersection of the
two spaces is null and there is no crossing items between the two
matrixes. Comparably if one space is the subspace or equivalent to
the other space, the overlap then is nontrivial generally since the
two matrixes belong to the same space. Hence in order to find the
maximal overlap between two matrixes, it is also \emph{necessary}
for the two matrixes to be in the same space. From physical point,
it means that the two operator is \emph{necessary} commutative.
Furthermore it is easy to understand why the maximal GE for pure
entangled state always happens for purely separable state.

As for the determination of $\rho^s$ in Eq.\eqref{delta}, it is
required for $\rho^s$ to be commutative to $\rho$, i.e. $[\rho,
\rho^s]=0$, which means that  $\rho^s$ shares the same global
symmetry with $\rho$. With this point one can determine $\varrho$ as
Eq.\eqref{srho}. Furthermore since $\rho^s$ is diagonal under the
collective basis $\{\ket{n-\tfrac{N}{2}}, n=0,1,\cdots,N\}$,
\begin{equation}
\text{Tr}[\rho\rho^s]=\sum_{n}\rho_{nn}\rho^{s}_{nn}\leq\sum_{n}\frac{\rho_{nn}^2+(\rho^{s}_{nn})^2}{2}.
\end{equation}
where $\rho_{nn}$ and $\rho^{s}_{nn}$ denote the diagonal elements
of $\rho$ and $\rho^s$ respectively. Obviously the second equality
occurs if and only if $\rho_{nn}=\rho^{s}_{nn}$, which means that
$\ex{J_z}$ has same value for both $\rho$ and $\rho^s$. And then $a$
can be determined in Eq.\eqref{srho}.

As for the state
\begin{equation}
\ket{\psi}=\frac{1}{\sqrt{2}}(\ket{1010}+\ket{0101}),
\end{equation}
our discussion above is also applicable. We should emphasize our
point clearly in this place that  $\ket{\psi}$ \emph{is not really
translational invariance}. Actually when one talks of the
translational invariance about a systems, it means
\begin{equation}
DHD^{\dagger}=H,
\end{equation}
in which $D$ is the translation operator, and $H$ is the Hamiltonian
for this system. Hence that one speaks of the translational
invariance for a state is meaningless without specifying the
Hamiltonian. Our discussion about Dicke model manifests clearly this
point. So the crucial point is to find the Hamiltonian for which
$\psi$ is the eigenvector. It seems that one can construct the
following Hamiltonian
\begin{equation}\label{h1}
H=\sum_{i}\sigma_i^z\sigma_{i+1}^z,
\end{equation}
for which $\psi$ is seemingly one of the degenerate ground states.
If the translational invariance is required for this system, one
must have the periodic boundary condition
$\sigma_{N+1}^z=\sigma_1^z$, where $N$ is the total particle number.
However $\ket{\psi}$ tells us that for one particle, its neighbored
particles always has opposite state to its state, which obviously
does not satisfy this periodic boundary condition. So we argue in
this place that the translational invariance for $\ket{\psi}$ is
only occasional because of its special form.

Instead $\ket{\psi}$ is the true ground state for Hamiltonian
\begin{equation}\label{h2}
H=-\sum_{i}\sigma_i^z\sigma_{i+2}^z,
\end{equation}
with the boundary condition $\sigma_{N+2}^z=\sigma_2^z$. It means
that the particle always has the same state to its next neighbored
particle. Thus it could explain naturally the reason that the
maximal overlap with $\ket{\psi}$ happens for the fully separable
states $\ket{1010}$ or $\ket{0101}$, which obviously satisfy this
boundary condition and also are the ground states for this
Hamiltonian.

In another point, one can also find a state which seemingly
satisfies the requirement of the "translational invariance" defined
by $\ket{\psi}$, i.e.,
\begin{equation}
\rho'=\frac{1}{2}(\ket{0101}\bra{0101}+\ket{1010}\bra{1010}),
\end{equation}
which obviously maximize the overlap with $\ket{\psi}$. This
features demonstrate again that $\ket{\psi}$ is not truly
translational invariance since $\rho'$ is the incoherent
superposition of the two degenerate ground states for Eq.\eqref{h2}.

\subsection*{Appendix B: Derivation of Eq.\eqref{toverlap}}
Set
\begin{eqnarray}
H_0&=&\omega a^{\dagger}a;\nonumber\\
H_I&=&\omega_0J_z+\frac{2\lambda}{\sqrt{N}}(a^{\dagger}+a).
\end{eqnarray}
Under $\beta=\tfrac{1}{k_BT}\ll1$, the partition function can be
approximated as\cite{lz05}
\begin{eqnarray}
z&=&\text{Tr}[e^{-\beta(H_0+H_I)}]\nonumber\\
&=&\text{Tr}[e^{-\beta H_0/2}e^{-\beta H_I/2}e^{-\beta
H_0/2}+O(\beta^3)]\nonumber\\
&\simeq&\text{Tr}[e^{-\beta H_0}e^{-\beta H_I}].
\end{eqnarray}
With respect of
\begin{equation}
\rho^s=\sum_{n=1}^{N}{_N C_n}a^n
(1-a)^{N-n}\ket{n-\tfrac{N}{2}}\bra{n-\tfrac{N}{2}},
\end{equation}
then
\begin{eqnarray}
\Delta&=&\frac{1}{z}\text{Tr}[\rho\rho^s]\nonumber\\&=&\frac{1}{z}\text{Tr}[\sum_{k=1}^{N}{_N
C_n}a^n (1-a)^{N-n}\nonumber\\&&\bra{n-\tfrac{N}{2}}e^{-\beta
H_0}e^{-\beta H_I}\ket{n-\tfrac{N}{2}}].
\end{eqnarray}
for which $[\rho^s, H_0]=0$ is applied. The tricky for the tracing
in the equation above is noting that
$\ket{\tfrac{N}{2};n-\tfrac{N}{2}}$ denotes the state in which $n$
particles are spin-up, and the others are spin-down. And then
\begin{eqnarray}
&&\bra{n-\tfrac{N}{2}}e^{-\beta H_0}e^{-\beta
H_I}\ket{n-\tfrac{N}{2}}\nonumber\\
&=&e^{-\beta \omega a^{\dagger}a}\bra{n-\tfrac{N}{2}}\prod_{i=1}^N
e^{-\beta[\tfrac{\omega_0}{2}\sigma_i^z+\tfrac{\lambda}{\sqrt{N}}(a^{\dagger}+a)\sigma_i^x]}
\ket{n-\tfrac{N}{2}}\nonumber\\
&=&e^{-\beta \omega a^{\dagger}a}\bra{n-\tfrac{N}{2}}\otimes_{i=1}^N
\sum_{k=0}^{\infty}
\frac{\beta^{2k}}{(2k)!}[\frac{\omega_0^2}{4}+\frac{\lambda^2}{N}(a^{\dagger}+a)^2]^k\nonumber\\
&&\{1-\frac{\beta}{2k+1}[\frac{\omega_0}{2}\sigma_i^z+\frac{\lambda}{\sqrt{N}}(a^{\dagger}+a)\sigma_i^x]\}
\ket{n-\tfrac{N}{2}}\nonumber\\
&=&e^{-\beta \omega a^{\dagger}a}\left\{\sum_{k=0}^{\infty}
\frac{\beta^{2k}}{(2k)!}[\frac{\omega_0^2}{4}+\frac{\lambda^2}{N}(a^{\dagger}+a)^2]^k(1-\frac{\beta}{2k+1}\frac{\omega_0}{2})\right\}^n\nonumber\\
&&\left\{\sum_{k=0}^{\infty}
\frac{\beta^{2k}}{(2k)!}[\frac{\omega_0^2}{4}+\frac{\lambda^2}{N}(a^{\dagger}+a)^2]^k(1+\frac{\beta}{2k+1}\frac{\omega_0}{2})\right\}^{N-n}
\end{eqnarray}
Thus
\begin{eqnarray}\label{1}
\Delta=\frac{1}{z}\text{Tr}[e^{-\beta \omega
a^{\dagger}a}&&\left\{\sum_{k=0}^{\infty}
\frac{\beta^{2k}}{(2k)!}[\frac{\omega_0^2}{4}+\frac{\lambda^2}{N}(a^{\dagger}+a)^2]^k \right. \nonumber \\
&&\left. [1+\frac{\beta}{2k+1} \frac{\omega_0}{2} (1-2a)]\right\}^N]
\end{eqnarray}
Expand the item in the corbeil bracket
\begin{eqnarray}\label{2}
\Rightarrow&&\sum_{k_1=0;k_2=0\cdots
k_N=0}^{\infty}\left(\prod_{i=1}^N
\frac{\beta^{2k_i}}{(2k_i)!}[1+\frac{\beta}{2k+1} \frac{\omega_0}{2}
(1-2a)]\right)\times\nonumber\\
&&\sum_{q=0}^{K=k_1+k_2+\cdots+k_N}\frac{K!}{q!(K-q)!}(\frac{\omega_0}{2})^{2(K-q)}
(\frac{\lambda}{\sqrt{N}})^{2q}(a^{\dagger}+a)^{2q}.
\end{eqnarray}
Define $a^{\dagger}a\ket{m}=m\ket{m}$, and then
\begin{eqnarray}
\Delta&=&\frac{1}{z}\sum_{k_1=0;k_2=0\cdots
k_N=0}^{\infty}\left(\prod_{i=1}^N
\frac{\beta^{2k_i}}{(2k_i)!}[1+\frac{\beta}{2k+1} \frac{\omega_0}{2}
(1-2a)]\right)\nonumber\\
&&\times
\sum_{q=0}^{K=k_1+k_2+\cdots+k_N}\frac{K!}{q!(K-q)!}(\frac{\omega_0}{2})^{2(K-q)}
(\frac{\lambda}{\sqrt{N}})^{2q}\nonumber\\&&\left.\frac{\text{d}^{2q}}{\text{d}
\eta^{2q}}e^{\tfrac{\eta^2}{2}}\sum_{m=0}^{\infty}
e^{-m\beta\omega}L_m(-\eta^2)\right|_{\eta=0},
\end{eqnarray}
where $L_m(x)$ is the $m$th Laguerre polynomial, and the relation is
used
\begin{eqnarray}
\bra{m}(a^{\dagger}+a)^{2q}\ket{m}&=&\left.\frac{\text{d}^{2q}}{\text{d}
\eta^{2q}}\bra{m}e^{\eta(a^{\dagger}+a)}\ket{m}\right|_{\eta=0}\nonumber\\
&=&\left.\frac{\text{d}^{2q}}{\text{d}
\eta^{2q}}[e^{\tfrac{\eta^2}{2}}L_m(-\eta^2)]\right|_{\eta=0}.
\end{eqnarray}

Apply the relation
\begin{eqnarray}
\sum_{m=0}^{\infty}
e^{-m\beta\omega}L_m(-\eta^2)=\frac{1}{1-e^{-\beta\omega}}\exp[\eta^2\frac{1}{e^{\beta\omega}-1}],
\end{eqnarray}
and then
\begin{eqnarray}
\Delta&=&\frac{1}{z}\frac{1}{1-e^{-\beta\omega}}\sum_{k_1=0;k_2=0\cdots
k_N=0}^{\infty}\nonumber\\&&\left(\prod_{i=1}^N
\frac{\beta^{2k_i}}{(2k_i)!}[1+\frac{\beta}{2k+1} \frac{\omega_0}{2}
(1-2a)]\right)\nonumber\\
&&\times
\sum_{q=0}^{K=k_1+k_2+\cdots+k_N}\frac{K!}{q!(K-q)!}(\frac{\omega_0}{2})^{2(K-q)}
(\frac{\lambda}{\sqrt{N}})^{2q}\nonumber\\&&\left.\frac{\text{d}^{2q}}{\text{d}
\eta^{2q}}e^{\tfrac{\eta^2}{2}\coth\tfrac{\beta\omega}{2}}\right|_{\eta=0}.
\end{eqnarray}
With the relations
\begin{eqnarray}
\left.\frac{\text{d}^{2q}}{\text{d}
\eta^{2q}}e^{\tfrac{\eta^2}{2}\coth\tfrac{\beta\omega}{2}}\right|_{\eta=0}
=(2q-1)!!\coth^q\tfrac{\beta\omega}{2}\nonumber\\
(2q-1)!!=\sqrt{\frac{A}{\pi}}2^p
A^p\int_{-\infty}^{\infty}\text{d}xe^{-Ax^2}x^{2p},
\end{eqnarray}
and set $A=1/2$,
\begin{eqnarray}
\Delta&=&\frac{1}{z}\frac{\sqrt{1/2\pi}}{1-e^{-\beta\omega}}\int_{-\infty}^{\infty}\text{d}xe^{-x^2/2}\sum_{k_1=0;k_2=0\cdots
k_N=0}^{\infty}\nonumber\\&&\left(\prod_{i=1}^N
\frac{\beta^{2k_i}}{(2k_i)!}[1+\frac{\beta}{2k+1} \frac{\omega_0}{2}
(1-2a)]\right)\times\nonumber\\
&&
\sum_{q=0}^{K=k_1+k_2+\cdots+k_N}\frac{K!}{q!(K-q)!}(\frac{\omega_0}{2})^{2(K-q)}
(\frac{x^2\lambda^2}{N}\coth\tfrac{\beta\omega}{2})^{q}.
\end{eqnarray}
Finally inverse the procedure from Eq.\eqref{1} to Eq.\eqref{2} for
the sum item and apply relations $\cosh x=\tfrac{e^x+e^{-x}}{2}$ and
$\sinh x=\tfrac{e^x-e^{-x}}{2}$, one then obtains the
Eq.\eqref{toverlap}.

\end{document}